\documentclass[aps,prb,twocolumn,superscriptaddress,floatfix]{revtex4-2}

\usepackage{graphicx} 
\usepackage{appendix}
\usepackage{amsmath, amssymb, braket}
\usepackage{dirtytalk}
\usepackage{hyperref, listings}
\usepackage{bbold}
\usepackage{esvect, cancel}
\usepackage{svg}
\usepackage{comment}
\usepackage{float}

\begin{document}

\title{Probing fluctuations in photon emission statistics of spin injected quantum dots with time dependent electronic excitation}

\author{Chiran Wijesundara}
\email{chiranya@buffalo.edu}
\affiliation{Department of Physics, University at Buffalo SUNY, Buffalo, New York 14260, USA}
\author{Vincent Iglesias-Cardinale}
\affiliation{Department of Physics, University at Buffalo SUNY, Buffalo, New York 14260, USA}
\author{Yuan Lu}
\affiliation{Institut Jean Lamour, Université de Lorraine, CNRS, UMR 7198, 54011 Nancy, France}
\author{Herbert Fotso}
\affiliation{Department of Physics, University at Buffalo SUNY, Buffalo, New York 14260, USA}
\author{Bo Xu}
\affiliation{Key Laboratory of Semiconductor Materials Science, Institute of Semiconductors, Chinese Academy of Sciences,P.O. Box 912, Beijing 100083, China}
\author{Tim Thomay}
\email{thomay@buffalo.edu}
\affiliation{Department of Physics, University at Buffalo SUNY, Buffalo, New York 14260, USA}

\begin{abstract}
\noindent {Electrically spin-injected ensembles of InGaAs/GaAs quantum dots emit circularly polarized single photons, with the emission helicity determined by the spin polarization of the injected electrons. A ferromagnetic CoFeB spin-injection layer with perpendicular magnetic anisotropy, combined with an MgO tunnel barrier, enables spin-polarized electron injection into the quantum-dot active region, resulting in  circularly polarized photon emission at zero applied magnetic field. These devices realize direct spin-charge-photon conversion in an electrically driven, microelectronics-compatible quantum light source that operates without an external pump laser. Using an extended Hanbury Brown–Twiss interferometer with four detectors and time-tagged readout, we measure time-resolved photon-number statistics that resolve up to $|2\rangle$ Fock states within a single excitation cycle. While single-photon ($n=|1\rangle$) emission tracks the electrical excitation pulse as expected, the two-photon ($n=|2\rangle$) statistics exhibit pronounced fluctuations at multiple temporal positions within the pulse. These fluctuations vanish upon demagnetization of the spin-injection layer, establishing a direct link between the multi-photon emission dynamics and the spin polarization of the injected carriers. The observations indicate that spin injection controls not only the helicity of the emitted photons but also higher-order temporal correlations of the emission. The behavior is phenomenologically captured by inserting both a variable excitation mechanism and a set of multi-emitter collapse operators within a stochastic wave function evolution. The platform thus opens a route to electrically driven, spin-controlled single- and multi-photon Fock-state sources and provides a testbed for spin-conditioned multi-emitter dynamics in solid-state quantum light sources.}
\end{abstract}

\maketitle


\section{Introduction}
\label{sec:Introduction}

\begin{figure*}[t]
    \centering
    \includegraphics[width=\textwidth]{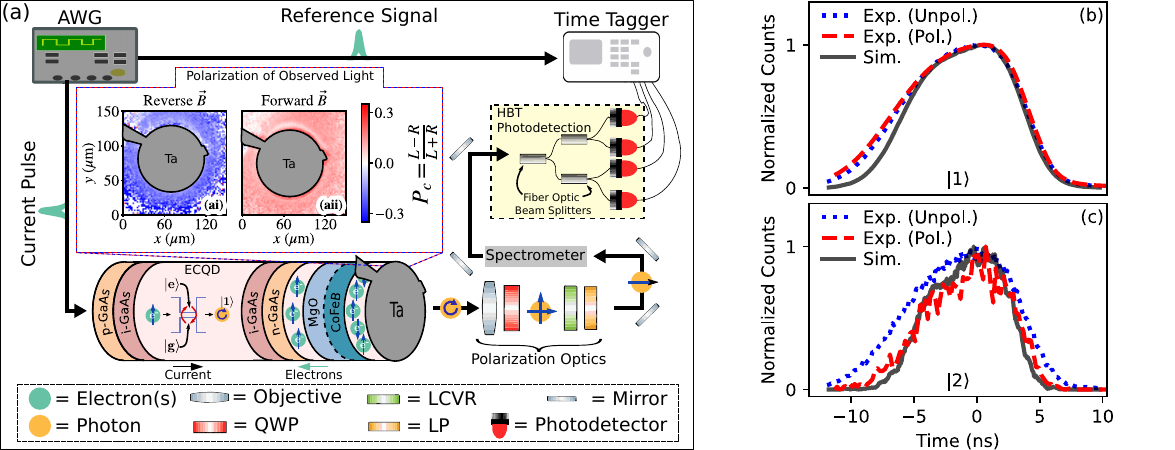}
    \caption{(a) \textbf{Experimental set up:} Beginning with the arbitrary waveform generator (AWG), an electronic pulse is applied to the ECQD ensemble depicted here in its place in the cylindrical sample. A reference pulse is simultaneously sent to the time tagger. The excitation stimulates the emission of photons into a full $4\pi$ solid angle which are collected using an infinity-corrected objective located coaxially with the stack and analyzed with a series of optical elements. Polarization analysis is performed using a quarter-wave plate (QWP) in combination with a liquid-crystal variable retarder (LCVR) and a linear polarizer (LP) to isolate spin components. The signal is then spectrally resolved by a grating spectrometer and subsequently coupled into a so called extended fiber-based Hanbury Brown–Twiss interferometer (highlighted in yellow) for correlation measurements carried out using fiber-optic beam splitters (FBSs) and avalanche photodiodes (APDs). Finally, the time tagger records the arrival times of all detected photons relative to the initial electrical trigger, enabling time-resolved correlation analysis. A polarization bias in emitted photons can be achieved via application of a magnetic field along the growth direction of the cylindrical sample. This is highlighted in panels (ai) and (aii). (b) \textbf{Time-resolved results:} Counts of $|1\rangle$- and $|2\rangle$-photon events as a function of time collected by the time tagger. Both are normalized to unity for direct comparison.}
    \label{fig:experiment}
\end{figure*}

\noindent Photo-excited quantum dots (QDs) have long demonstrated their potential as robust single-photon emitters~\cite{sotier_femtosecond_2009,santori_indistinguishable_2002,senellart_high-performance_2017,sapienza_nanoscale_2015,zhao_single_2018,arakawa_progress_2020}. Traditionally, the classical optical properties of QD emission have been exploited in applications such as optical telecommunications and three-dimensional display technologies~\cite{cadiz_electrical_2018,dainone_controlling_2024,lu_spin_2026}. While incorporating quantum light sources directly into microelectronic platforms is desirable for a number of purposes including device miniaturization, operational efficiency, and scalability~\cite{guo_electrically_2023, yuan_electrically_2002, lin_electrically_2019, mizuochi_electrically_2012}, it remains an elusive target as most single-photon sources reported to date rely on optical excitation. Electrically contacted quantum dots (ECQDs), or optical two-level emitters combined with an electron-hole excitation mechanism, represent a promising route toward such an implementation.

The device studied in this work~\cite{giba_spin_2020} consists of ECQDs capable of efficient spin injection and subsequent conversion of carrier spin polarization into circularly polarized light~\cite{cadiz_electrical_2018}. In contrast to conventional photo-excited QDs, electrically excited quantum dots exhibit more complex time dynamics, owing to the stochastic nature of electrical carrier injection. Applying a variable current to such an ensemble of ECQDs allows for tunable excitation, and thereby tunable photon emission intensity. Indeed, even if the precise interaction between the bulk current and the individual emitters is not known, measurement of the emitted light can serve as an indicator of the internal dynamics since each emitted photon corresponds to the recombination of an injected charge carrier.

Spin injected ECQDs generally comprise two functional regions: a ferromagnetic layer responsible for injecting spin-polarized carriers, and a semiconductor region in which exciton recombination results in emission of circularly polarized photons~\cite{karli_controlling_2024,prezhdo_photoinduced_2009,truong_high_2009}. Electrical excitation offers additional tunability compared to optical pumping, including control over repetition rate, pulse-rise and pulse-fall times, and overall pulse duration. These parameters directly influence the exciton population dynamics and thus the photon emission statistics.
The characterization of quantum light sources relies on identifying sub-Poissonian photon-number distributions, in which the probability of detecting 
$\ket{n}$-photons in a given interval demonstrates antibunching, a hallmark of non-classical light~\cite{fox_quantum_2006,paul_photon_1982,loudon_photon_1976}. The performance of a single-photon source is therefore commonly assessed in terms of its ability to suppress multi-photon emission events. Nevertheless, higher-order photon emissions do occasionally occur, and rather than being merely detrimental, they can be exploited to optimize single-photon performance or to engineer reliable sources of higher-order Fock states.
Recent advances in photonic quantum technologies including quantum communication, computation, and quantum-enhanced metrology have highlighted the importance of higher-order photon Fock states~\cite{holland_interferometric_1993,nagata_beating_2007,giovannetti_advances_2011,yamamoto_preparation_1986}. Understanding the internal dynamics of quantum emitters such as ECQDs is therefore essential for advancing both single-photon and higher-order photon-state generation.

In this paper, we present an experimental demonstration of non-trivial fluctuations of $|2\rangle$-photon events in the time domain when observing emissions from an electronically excited quantum dot source. This is along with a simulation scheme that phenomenologically captures similar behavior for similar event counts. The main experimental results are found in Section~\ref{sec:Experiment} which also contains a brief overview of the experimental procedure. An explanation of the simulation is given in Section~\ref{sec:Methods_sim} which includes also the combined \textit{Experiment + Simulation} results. Discussion of those results including a noise analysis is presented in Section~\ref{sec:discussion} before a summary of the findings is finally given in Section~\ref{sec:conclusion}. Further details regarding particulars of the experimental methods are reserved for Section~\ref{sec:methods}.
\section{Experiment results}
\label{sec:Experiment} 

The experimental setup can be organized into three stages based on functionality. 
The excitation stage, the photon collection and polarization manipulation stage and the higher order detection stage.
A schematic diagram of the excitation stage is labeled 'ECQD' in the depiction of the cylindrical sample in Fig.~\ref{fig:experiment}(a). 
The device is housed in a cryostat with an optical port as a sub $100~K$ temperature is required for it to function.
The arbitrary waveform generator (AWG)  produces a sequence of tunable voltage pulses that act as the excitation mechanism for the ECQDs and provide an experimental advantage over optical excitation via the versatility in pulse control.

The QDs subsequently emit single photons into a full $4\pi$ solid angle, though photon collection is done in free space via an infinity corrected objective placed coaxially with the sample. The photons' polarization states are determined by the injected electron spin. 
This is a direct result of the $X^+$ trion spin system. As the ground state forms a spin singlet, the only contribution towards the average spin is from the injected electron. 
Coupled with the Fermi selection rules this results in the emission of circularly polarized single photons from each two level system. 
The control of the average spin injected into the QDs is facilitated by the perpendicular magnetic anisotropy (PMA) present in the CoFeB layer (even at zero external field)~\cite{tao_atomic-scale_2018,liang_large_2014}. Figures \ref{fig:experiment} (ai) and (aii) show a mapping of the spatial distribution of $P_C$, the left/right circular polarization bias under opposite magnetization directions. The polarization is distributed uniformly across the sample, indicating that spin injection is not localized to a small subset of quantum dots, a very desirable quality for a spin polarized photon source. See Section~\ref{sec:spatial_map} for details regarding this measurement. 

Demagnetizing the sample allows for isolation of the role of spin injection in the photon statistics. To demagnetize it, the sample was rotated at high speed in a strong magnetic field while the field was ramped down, which breaks the film into many small domains with randomly oriented magnetization. The net magnetization therefore averages to zero, resulting in zero net spin injection, thus removing the polarization bias from emitted photons. Everything else about the device and the excitation is left unchanged. Throughout the remainder of the paper, "unpolarized" refers to data taken after this demagnetization procedure, and "polarized" to data taken before it, with spin injection active.



Circularly polarized emissions from the quantum dots are converted to linear polarization using a quarter-wave plate. 
The resulting linear light is then passed through a liquid-crystal variable retarder (LCVR) acting as a half wave-plate and is subsequently sent through a linear polarizer to select either the left or right handed polarization component.
After the polarization optics, the light is directed into a spectrometer, which isolates the desired emission wavelength. 
The spectrometer enables both zero order imaging with a CMOS camera and the spectral filtering of the photons that are routed into the detection stage.

$g^{(2)}(0)$ is understood to be a number that reflects the average photon statistics over all the dynamics that happens during the pulse. This deficiency is particularly exaggerated when the excitation pulse exceeds the lifetime of the emitter. It is expected of a single quantum emitter to behave as an anti-bunched source, however an ensemble suffering an excitation pulse exceeding its lifetime behaves similar to a coherent source~\cite{carmichael_intensity_1978,kimble_multiatom_1978}. 
This warrants further investigation as a simple $g^{(2)}(0)$ value, even when measured with time resolution, cannot fully capture the underlying photon emission dynamics.

Here, we employ an extended fiber-based Hanbury Brown and Twiss (HBT) setup \cite{brown_correlation_1956} that expands detection beyond standard $g^{(2)}(\tau)$ by capturing these higher-order photon statistics.
The setup extends the traditional single 50:50
beam splitter and two single photon detectors (in this case avalanche photo detectors (APDs)) to three beam splitters and four detectors. 
To distinguish $\ket{n}$-photon higher-order states a minimum of $n+1$ detectors are required. With four APDs, correlations up to the $\ket{3}$-photon state can be resolved. 
The time tagger is operated in the Time-Tagged Time-Resolved (TTTR) data acquisition mode. The time-tagging module records the photon arrival times at each detector relative to the reference trigger pulse generated by the AWG.
The setup is based on our previous publication~\cite{thomay_simultaneous_2017,xu_optimized_2024,carvalho_spectrally-resolved_2025} to explore time resolved photon statistics.

The HBT setup enables characterization of the time resolved photon statistics of the emitted light, most prominently allowing us to distinguish single-photon emission from higher-order photon states.
With a coincidence window (see Section \ref{subsec:hbt}) we are able to reconstruct the photon statistics seen in Fig.~\ref{fig:experiment} (b) and (c).
For an electrical excitation of $20~ns$ pulse width prepared with the AWG, the $|1\rangle$ photon emissions follow the curves presented in panel (b) and the $|2\rangle$ photon emissions follow the curves presented in panel (c).
Due to capacitive and resistive elements present before the electrical pulse reach the quantum dots, it is difficult to predict the precise excitation statistics of the ensemble ECQDs however it is clear that the response is different between the $|1\rangle$- and $|2\rangle$-photon statistics. 
Pronounced fluctuations are observed in the $\ket{n=2}$ Fock-state population.

\begin{figure}[t]
    \centering
    \includegraphics[width=\columnwidth]{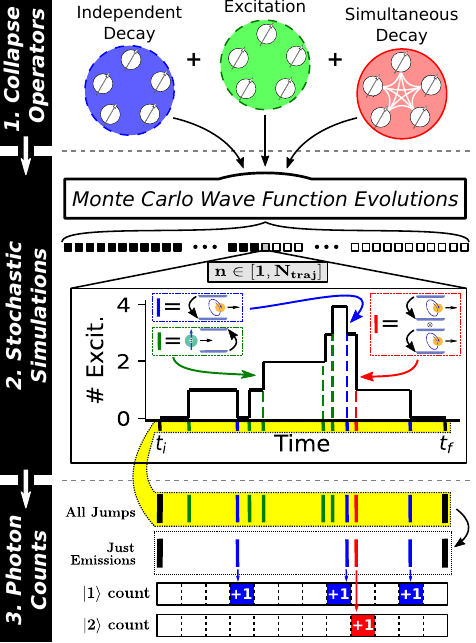}
    \caption{The simulation scheme begins with constructing the desired collapse operators. These are then implemented in a Monte Carlo wavefunction method, stochastically producing jump times. Excitations are disregarded and the $|1\rangle$- and $|2\rangle$-photon events are identified based on the jump times of their associated operator.}
    \label{fig:sim_schematic}
\end{figure}

To elucidate the underlying mechanism, we turn to theoretical modeling of coherent dynamics in an ensemble of emitters.

\begin{figure*}[t]
    \centering
    \includegraphics[width=7in]{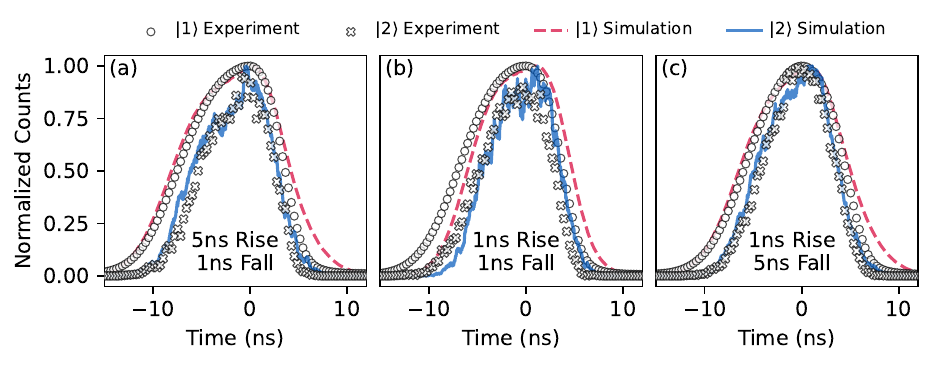}
    \caption{Variation of the applied current pulse, described in each panel by the rising and falling time of the pulse, produces mild variations in the experimentally observed $|1\rangle$-photon events. Constructing the excitation function $\Omega(t)$ so the simulated $|1\rangle$-photon curve matches the corresponding experimental data produces significant overlap in the simulated and experimental $|2\rangle$-photon curves.}
    \label{fig:shapes}
\end{figure*}
\section{Theoretical model}
\label{sec:Methods_sim}

We model the ensemble of ECQDs phenomenologically by inserting various sets of collapse pathways into a Monte-Carlo Wavefunction evolution~\cite{Plenio_Knight_Review, Dalibard_etal, Molmer:93}. Figure~\ref{fig:sim_schematic} elucidates this process through which the experiment is simulated using the full set of collapse (and excitation) operators. 

The first set of collapse operators arises from the master-equation treatment of the many-emitter Weisskopf-Wigner Hamiltonian,

\begin{equation}
\begin{aligned}
        H &= \frac{\hbar\omega_0}{2}\sum_i\sigma_i^z + \hbar\sum_{\mathbf k\lambda}\omega_{\mathbf k}a_{\mathbf k\lambda}^\dagger a_{\mathbf k\lambda}\\
        &\quad-i\hbar\sum_i\sum_{\mathbf k\lambda}g_{\mathbf k\lambda}^i(\sigma_i^+ + \sigma_i^-)(a_{\mathbf k\lambda}^\dagger - a_{\mathbf k\lambda} ),
    \end{aligned}
    \label{eqn:weisskopf-wigner+current}
\end{equation}

\noindent in which $\sigma_i^z=|e_i\rangle\langle e_i|-|g_i\rangle\langle g_i|$ and $\sigma_i^+ = (\sigma_i^-)^\dagger = |e_i\rangle\langle g_i|$ are the Pauli$-z$ and the raising/lowering operators for the two-level atom $i$ with energy gap $E_{0}=E_e-E_g=\hbar\omega_0$ between $|g_{i}\rangle$ its ground and $ |e_{i}\rangle$ its excited state, assumed here to be identical for all emitters. $a_{\mathbf k\lambda}$ ($a_{\mathbf k\lambda}^\dagger$) is the annihilation (creation) operator for the bosonic mode with wavevector $\mathbf k$ and polarization $\lambda$ which is coupled to atom $i$ with coupling strength $g_{\mathbf k\lambda}^i$. The Hamiltonian in this form is written in the long-wavelength limit~\cite{constantin2025}, so the coupling strength loses its dependence on the atomic index. Additionally, we assume all emitters and bosonic modes are polarized along a common axis. In a standard treatment, this is a crucial approximation for the establishment of cooperative emission.

When the atomic system is treated as a non-interacting open quantum system coupled to a continuum of bosonic modes, the emission process is found to be approximately Markovian and irreversible. The exponential decay is captured by the set of collapse operators

\begin{equation}
    J_{\text{ind}}=\{\sqrt{\Gamma_0}\sigma_1^-,\sqrt{\Gamma_0}\sigma_2^-,\cdots,\sqrt{\Gamma_0}\sigma_N^-\} 
    \label{eqn:ind-collapse}
\end{equation}

\noindent where $\Gamma_0$ determines the natural lifetime of independent emitters ($\tau = 1/\Gamma_0 = 800~ps$ in this case)~\cite{Cohen_Tannoudji_Book1992, Breuer:2002pc}. These operators correspond to the collapse of unique emitters in the ensemble, and thus the action of these operators corresponds to the emission of single photon from the ensemble. In this way, the jump times of these operators in the stochastic simulations are associated with the emission times of $|1\rangle$-photon events. 

Similarly, $|2\rangle$-photon events in this simulation method are considered to coincide with the simultaneous collapse of any pair of atoms, $|3\rangle$-photon events with the simultaneous collapse of any trio, and so on. Multiple $|1\rangle$-photon events coincident within a small time window, registered as a single higher order photon event experimentally, are registered in this simulation as independent $|1\rangle$-photon events. We hypothesize that spin injection enables cooperative dynamics, thus enabling the emission of higher order photon states via a novel collapse pathway. To account for these higher order photon events, we consider the dynamics of the system with jump operators of the form

\begin{equation}
    J_{\text{joint}}=\{\sqrt{\kappa}\sigma_i^-\otimes\sigma_j^-~|~i\ne j\in1,2,\cdots N\},
    \label{eqn:joint-collapse}
\end{equation}

\noindent where $\kappa$ is a tunable parameter that must in general account for the rate of $|2\rangle$-photon emissions and additionally in this implementation for the relative loss between $|1\rangle$-photon observations and $|2\rangle$-photon observations. In the results shown here, we take $\kappa = 0.0016$, chosen to best match the experimental ratio of the number of $|1\rangle$-photon events to the number of $|2\rangle$-photon events. The jump times of these operators are associated with the times of $|2\rangle$-photon emission events.

It should be noted that the joint collapse operators~(\ref{eqn:joint-collapse}) do not arise from a standard treatment of the Weisskopf-Wigner Hamiltonian nor are they typical of any decay channels known to the authors. Further consideration of a mechanism responsible for these collapses is a promising direction for future work, however a few notes can be made about these operators here. They are chosen because they ensure simultaneous collapse between atom pair $i$-$j$. The action of this collapse pathway is thus limited to the times during which two or more emitters are excited. Therefore, the excitation pulse profile and the rate of $|1\rangle$-photon emissions have direct impact on the $|2\rangle$-photon emission profile.

The last set of collapse operators utilized here is associated with the excitation. The applied current is a bulk macroscopic quantity, the interaction of which with emitters is unknown. Without exact microscopic specification, we model an effective excitation with the time dependent jump operators

\begin{equation}
    J_{\text{excite}}=\{\sqrt{\Omega(t)} \sigma_1^+,\sqrt{\Omega(t)}\sigma_2^+,\cdots,\sqrt{\Omega(t)} \sigma_N^+\}
    \label{eqn:excite-collapse}
\end{equation}

\noindent that excite atom $i\in[1,2,\cdots, N]$ from $|g\rangle$ to $|e\rangle$, weighted by $\Omega(t)$. The weighting function serves to model the current pulse by introducing a time dependent excitation probability for the ensemble. In alignment with the permutation symmetry of the Weisskopf-Wigner Hamiltonian, we assume that this weighting is equivalent for all atoms regardless of atomic index. The particular shape of $\Omega(t)$ is determined experimentally by $|1\rangle$-photon event statistics, as any emission must be preceded by an excitation, and the $|2\rangle$-photon counts are negligible in comparison. 

In each trial of the Monte-Carlo wavefunction evolution, we initialize three atoms in their ground states. Time evolution is then simulated using the combined set of all collapse operators $J_{\text{ind}}\cup J_{\text{joint}}\cup J_{\text{excite}}$. Excitation events signified by jump times of collapse operators in the set $J_{\text{excite}}$ are filtered out of the data, and photon emission events signified by jump times of the remaining collapse operators are stored. In this way we collect data comparable to that collected in the experiment and perform an identical analysis without the need to filter photon events based on their observation times. 


We choose a system size of $N=3$ spin-aligned emitters as a toy model that still captures the dynamics present in the experiment. Additionally, as mentioned above, cooperative emission requires a shared inter-emitter polarization axis. The experimental sample exhibits an average degree of polarization of $\sim21\%$ (see Figure~\ref{fig:experiment} (ai) and (aii)) and a sample density of $\sim$11 quantum dots per square emission wavelength ($904~nm$). Since the measured polarization directly reflects the spin population of the ensemble at the time of emission, these values reflect that fewer than 10 emitters are likely to engage in cooperative emission at any given time. This together with the unfavorable computational scaling of many-body quantum systems, justifies restricting the model to three emitters.

Strong deviation from the $|1\rangle$-photon event statistics was observed in the experimentally collected $|2\rangle$-photon event statistics. This behavior was reproduced within the simulation scheme, and is presented in Figure~\ref{fig:experiment} (b) and (c) as well as in Figure~\ref{fig:shapes}.

A dependence of the novel behavior on the polarization of the sample is confirmed in Figure~\ref{fig:experiment} (b) and (c) by comparing the time resolved $|2\rangle$-photon counts emitted by a polarized and an unpolarized sample. In the former case, presented in the dashed red line, the $|1\rangle$-photon events follow a smooth curve while the corresponding $|2\rangle$-photon events show pronounced deviations from that curve. In the latter case, presented in the dotted blue line, no such deviations are present. The solid black line presents data produced by the simulation scheme detailed above, and matches the results from the polarized sample.

Figure~\ref{fig:shapes} confirms the simulation captures the experimental $|2\rangle$-photon behavior over a range of incident current pulses. The current pulse (in each case with a total duration of $\sim$20ns) producing the experimental data has a rising and falling edge of 5ns and 1ns, respectively, in panel (a), 1ns and 1ns in panel (b), and 1ns and 5 ns in panel (c). These result in mildly different $|1\rangle$-photon statistics. To simulate the same current pulse, we construct $\Omega(t)$ in $J_{\text{excite}}$ so that the simulated $|1\rangle$-photon event curve qualitatively aligns with the experiment, then observe that the resulting $|2\rangle$-photon event curve also aligns with the experiment. Each data set presented here was produced with over 100,000 implementations of the Monte Carlo simulation. This many trials and the chosen value of $\kappa=0.0016$ result in $>10^7$ $|1\rangle$-photon events, and $>10^5$ $|2\rangle$-photon events. 



\section{Discussion}
\label{sec:discussion}

\begin{figure}[t]
    \centering
    \includegraphics[width=\columnwidth]{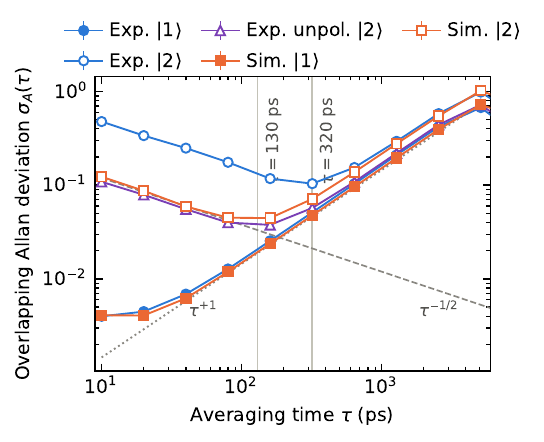}
    \caption{Comparison of Allan deviation of $|2\rangle$ and $|1\rangle$-photon for experimental (pol. and unpol.) and simulation. Reference lines (dotted and dashed) refer to standard mapping for physical noise sources. It should be noted the noise is in the time domain not in number state.}
    \label{fig:noise}
\end{figure}

Different emission mechanisms typically differ in their characteristic noise signatures. Thus a noise analysis provides further evidence for the proposed collapse mechanism proposed in the previous section.
Allan deviation was specifically selected because it serves as a metric for characterizing noise processes and mapping them to their underlying physical origins~\cite{allan_statistics_1966}.  In this study, we observe distinct statistical behaviors for $|1\rangle$-photon and $|2\rangle$-photon events, suggesting they are governed by separate noise mechanisms in shorter timescales. While a log-log gradient of $\alpha = -1/2$ is conventionally attributed to white phase or shot noise~\cite{allan_statistics_1966,huang_weak-value_2026}, the divergence in noise scaling between the $\ket{n=1}$ and $\ket{n=2}$ regimes is an unconventional phenomenon. This discrepancy provides a window into the difference in photon statistics and how we approach the matter of a quantum light source. Typically when classifying light sources one would expect noise, arising from the same source to follow the same statistics since the physical mechanism involved in the emission should be identical. However performing the error analysis after photon number resolving for the same source gives credence to treating the faster($320~ps$) $\ket{2}$-photon numbers to associate with a coherent process. It is further supported by the minima of unpolarized $\ket{2}$ having a smaller coherence time ($130~ps$). The simulation was also performed for two different lifetime measurements.



\section{Conclusion}
\label{sec:conclusion}
We examined the origin of the fluctuating emission observed exclusively in the $n=\ket{2}$ state in the spin injected case. Spin injection supplies the spin alignment that, together with the other requirements already met by the system, enables cooperative emission, and the fluctuations diminish in its absence. These observations suggest that cooperative effects are a plausible origin of the behavior. The microscopic mechanism that couples the emitters is not identified here, however the joint collapse operators in Eqn.~\ref{eqn:joint-collapse} capture its effect phenomenologically. Stochastic simulations are capable of reproducing the experimentally observed behavior, and, notably, show improved agreement when spin injection effects are present, further supporting the cooperative emission interpretation. Together, these results suggest that cooperative emission plays a significant role in the excited-state dynamics of ECQDs and should be considered in the design and interpretation of electrically driven single-photon sources.

\noindent \\

\section{Methods}
\label{sec:methods}
\subsection{Sample growth}
The spin light-emitting diode (spin-LED) consists of two parts: a semiconductor LED and a metallic spin injector.
The $p$-$i$-$n$ LED was grown by III-V molecular beam epitaxy (MBE) and contains a single layer of
In$_{0.3}$Ga$_{0.7}$As quantum dots (QDs) embedded in the active region.
The full layer sequence, from the substrate upward, was
$p^{+}$-GaAs:Zn (001) substrate ($p = 3\times10^{18}$~cm$^{-3}$),
300-nm $p$-GaAs:Be ($p = 5\times10^{18}$~cm$^{-3}$),
400-nm $p$-Al$_{0.3}$Ga$_{0.7}$As:Be
($p = 5\times10^{17}$--$5\times10^{18}$~cm$^{-3}$),
30-nm undoped ($i$) GaAs with a Be $\delta$-doping layer at its center,
7~monolayers (ML) of In$_{0.3}$Ga$_{0.7}$As forming the QDs,
30-nm $i$-GaAs, and
50-nm $n$-GaAs:Si ($n = 1\times10^{16}$~cm$^{-3}$).
The QDs had an areal density of $1.6\times10^{10}$~cm$^{-2}$, an average lateral diameter of
approximately 30~nm, and a height of approximately 9~nm.
The Be $\delta$-doping concentration was calibrated to yield approximately one hole per dot.
After growth, the LED surface was passivated with an arsenic capping layer in the MBE chamber, and the
structure was transferred through air to a second, interconnected MBE--sputtering system.
The As cap was desorbed at 300~$^{\circ}$C in the MBE chamber, and the sample was then transferred under
ultrahigh vacuum to the sputtering chamber for growth of the spin injector, which consisted of
2.5-nm MgO/1.1-nm Co$_{0.4}$Fe$_{0.4}$B$_{0.2}$/5-nm Ta.
Circular mesas 300~$\mu$m in diameter were defined by standard UV photolithography and etching.
Finally, the processed wafers were cleaved into small pieces and subjected to rapid thermal annealing (RTA)
at 300~$^{\circ}$C for 3~min to establish perpendicular magnetic anisotropy (PMA) in the spin injector.
Further details on the growth and optimization of the perpendicular spin injector are given in
Refs~\cite{liang_large_2014,tao_atomic-scale_2018}.

\subsection{Spatially resolved $P_c$ measurement}
\label{sec:spatial_map}
Polarization-mapping measurements were recorded using a CMOS-based imaging system. Each orthogonal spin component is selected by the LCVR acting as a half lambda plate. The use of an LCVR is essential because unlike a waveplate an LCVR does not introduce an image offset.  The selected polarization channel is further purified using a linear polarizer. The resulting signal is imaged in the zeroth diffraction order onto a CMOS sensor. 
To extract the circular polarization degree $P_c$, images corresponding to the two orthogonal polarization states are recorded and subsequently subtracted and normalized. 
The polarization mapping shows that the circular polarization degree $P_c$ remains spatially uniform. 
The polarization state reverses sign upon inversion of the external magnetic field, consistent with spin-injection mechanism.

\subsection{Time resolved HBT measurement}
\label{subsec:hbt}
The extended HBT setup used consisting of 4 single photon detectors is capable of resolving up to $n=\ket{3}$ photon numbers. However resolving $n=\ket{3}$ would require an impractically long acquisition time for this device, therefore only $n=\ket{1},\ket{2}$ were considered. The single-photon detectors (IDQ ID100) have a timing jitter of $40~\mathrm{ps}$, and the time-tagger (PicoQuant MultiHarp 150) has a jitter of $20~\mathrm{ps}$. The arbitrary waveform generator (Siglent SDG6022) operates at a repetition rate of $10^{7}~\mathrm{Hz}$ with a jitter of $100~\mathrm{ps}$. Coincidence histograms were binned at $10~\mathrm{ps}$, and a coincidence window of $500~\mathrm{ps}$ was used to identify $n=\ket{2}$ events. Each dataset was acquired with an integration time of $8~\mathrm{h}$. For the unpolarized sample, this yielded $5.1\times10^{8}$ $\ket{1}$ events and $3\times10^{5}$ $\ket{2}$ events; for the polarized sample, the corresponding counts were $3.8\times10^{8}$ and $7\times10^{4}$, respectively.
\noindent
\section{Acknowledgments}
Y.L. thank the support by the French National Research Agency (ANR) SOTspinLED projects (No. ANR-22-CE24-0006-01) and France 2030 government investment plan managed by ANR under grant PEPR SPIN—OptoSpinCom (No. ANR-24-EXSP-0010), and by the European Union’s Horizon Europe Research and Innovation program SpinDataCom under Grant Agreement number 101184694. Part of experiments were performed using equipment from the CC-DAUM platform funded by FEDER (EU), ANR, the Region Lorraine and the metropole of Grand Nancy.
\bibliography{bibliography}

\end{document}